\documentclass[a4]{article}
\usepackage[T1]{fontenc}
\usepackage[latin1]{inputenc}
\usepackage{amsmath, amssymb, amsthm}
\usepackage[english]{babel}
\usepackage{indentfirst}
\usepackage{array}
\usepackage{caption,appendix}
\usepackage{geometry}
\usepackage{float}
\usepackage{cancel}
\usepackage{bbold}
\usepackage{authblk}
\usepackage{cite}
\usepackage{color}
\numberwithin{equation}{section}

\theoremstyle{definition}

\begin{document}
\author[1,2,3]{Salvatore Capozziello \thanks{capozziello@na.infn.it}}
\author[4]{Maurizio Capriolo \thanks{mcapriolo@unisa.it} }
\author[4,5]{Gaetano Lambiase \thanks{glambiase@unisa.it}}
\affil[1]{\emph{Dipartimento di Fisica "E. Pancini", Universit\`a di Napoli {}``Federico II'', Compl. Univ. di
		   Monte S. Angelo, Edificio G, Via Cinthia, I-80126, Napoli, Italy, }}
\affil[2]{\emph{Istituto Nazionale di Fisica Nucleare (INFN), Sezione  di Napoli, Compl. Univ. di
		   Monte S. Angelo, Edificio G, Via Cinthia, I-80126,  Napoli, Italy,}}
\affil[3]{\emph{Scuola Superiore Meridionale, Largo S. Marcellino 10,  I-80138,  Napoli, Italy,}}
\affil[4]{\emph{Dipartimento di Fisica Universit\`a di Salerno, via Giovanni Paolo II, 132, Fisciano, SA I-84084, Italy.} }
\affil[5]{\emph{INFN, Gruppo Collegato di Salerno, via Giovanni Paolo II, 132, Fisciano, SA I-84084, Italy.} }
\date{\today}
\title{\textbf{The gravitational energy-momentum pseudo-tensor in $f(Q)$ non-metric gravity}}
\maketitle
\begin{abstract}
We derive the affine tensor associated with the energy and momentum densities of both  gravitational and  matter fields,  the complex pseudo-tensor, for $f(Q)$ non-metric gravity, the straightforward extension  of Symmetric Teleparallel Equivalent of General Relativity (STEGR),  characterized by  a flat, torsion-free, non-metric connection. The local conservation of  energy-momentum complex on-shell is satisfied through a continuity equation. An important analogy is pointed out between gravitational pseudo-tensor of  teleparallel $f(T)$ gravity,  in the Weitzenb\"ock gauge, and the same object of   symmetric teleparallel  $f(Q)$ gravity, in the coincident gauge.
Furthermore, we  perturb the gravitational pseudo-tensor $\tau^{\alpha}_{\phantom{\alpha}\lambda}$ in the coincident gauge up to the second order in the metric perturbation, obtaining a useful expression for  the power carried by the related gravitational waves. We also present an application of the gravitational pseudotensor, determining the gravitational energy density of a Schwarzschild spacetime in STEGR gravity, adopting the concident gauge. Finally,  analyzing the conserved quantities on manifolds, the Stokes theorem can be formulated for generic affine connections.
\end{abstract}
\section{Introduction}\label{sec1}
In General Relativity (GR), gravity is interpreted as the curvature of spacetime. This is because the affine connection is metric-compatible and symmetric. Such a connection  can be uniquely expressed in terms of the metric tensor $g_{\mu\nu}$. Hence, the Riemann tensor $R^{\alpha}_{\phantom{\alpha}\beta\mu\nu}$, which  is expressed in terms of the connection, can be reformulated as a function of the metric alone. Therefore, in GR, the curvature of the connection is non-zero while its torsion and non-metricity are zero. In conclusion, we can  say that if the connection gives rise only to curvature, it can be interpreted as a clue to the presence of gravity. 

For a generic non-metric connection with torsion and curvature, gravity can be interpreted as a property of  affine connection only, i.e. it is related to the curvature tensor $R^{\alpha}_{\phantom{\alpha}\beta\mu\nu}$, the torsion tensor $T^{\alpha}_{\phantom{\alpha}\mu\nu}$, and the non-metricity tensor $Q_{\alpha\mu\nu}$. On the other hand, in metric teleparallel theories, like $f(T)$ gravity, with $T$ the torsion scalar defined later, only the torsion of  connection is responsible for the gravitational interaction, even though, being metric, it can be expressed in a unique way in terms of the metric tensor. While in symmetric teleparallel theories, like $f(Q)$ gravity with $Q$ the non-metricity scalar defined later, where the connection is flat and symmetric, the gravitational field is encoded in  non-metricity. If the connection has no curvature, it is symmetric and metric-compatible: We are in  absence of gravity and the spacetime is  Minkowski. In summary, gravitational effects are due to the properties of  connection, and can be classified as follows:
\begin{itemize}
    \item {\bf Metric-affine geometry}: This is the most general case. $R^{\alpha}_{\phantom{\alpha}\beta\mu\nu}\neq 0$, $T^{\alpha}_{\phantom{\alpha}\mu\nu}\neq 0$ and $Q_{\alpha\mu\nu}\neq 0$, the gravity is encoded in curvature $R$, torsion $T$ and non-metricity $Q$ of connection;
    \item {\bf Riemann geometry}: This is   GR. We have $R^{\alpha}_{\phantom{\alpha}\beta\mu\nu}\neq 0$, $T^{\alpha}_{\phantom{\alpha}\mu\nu}=0$ and $Q_{\alpha\mu\nu}=0$, the gravity is encoded in curvature of connection $R$; 
    \item {\bf Metric teleparallel geometry}: As in Teleparallel Equivalent of GR (TEGR), $T^{\alpha}_{\phantom{\alpha}\mu\nu}\neq 0$, 
$R^{\alpha}_{\phantom{\alpha}\beta\mu\nu}=0$ and $Q_{\alpha\mu\nu}=0$,  the gravity is encoded in torsion of connection $T$;
    \item {\bf Symmetric teleparallel geometry}: As in Symmetric Teleparallel Equivalent of GR  (STEGR), $Q_{\alpha\mu\nu}\neq 0$, $R^{\alpha}_{\phantom{\alpha}\beta\mu\nu}=0$ and $T^{\alpha}_{\phantom{\alpha}\mu\nu}=0$, the gravity is encoded in non-metricity of connection $Q$.
    \item {\bf Minkowski geometry}: $R^{\alpha}_{\phantom{\alpha}\beta\mu\nu}=0$, $T^{\alpha}_{\phantom{\alpha}\mu\nu}=0$ and $Q_{\alpha\mu\nu}=0$, the gravitational interaction is absent and the spacetime is flat.
\end{itemize}
The connection is an affine and non-covariant object, and gravity is intimately connected to it. So if we calculate quantities such as energy or momentum  including the gravitational field, we obtain affine geometric objects. Furthermore, since the connection is defined on the whole manifold, the gravitational field also pervades the whole Universe and cannot be localized, that is, two massive bodies interact gravitationally at any distance. This property is a clear implication of the affine nature of gravitational energy-momentum. We might conclude that gravity is somehow an affine property of spacetime.

As  said above, if we turn off the curvature and torsion of  connection defined on the manifold which describes the spacetime, we obtain a non-metric connection.  Non-metricity can be considered also with non-linear extensions through  analytic functions as $f(Q)$~\cite{Heisenberg:2023lru,Gomes:2023tur}. Such theories that modify the geometry of spacetime without  introducing  exotic forms of matter and energy are becoming necessary since GR fails at  UV and IR scales and can be enclosed in the vast realm of extended and modified theories of gravity \cite{CANTATA:2021asi,Faraoni:2010pgm}. In fact, $f(Q)$ gravity can explain both the early  cosmological accelerations in the framework of inflationary paradigm \cite{Mehdi, Kurt, Sadatian, Chakra, Koivisto} and late-time acceleration without including dark energy  \cite{Nojiri, Enk, Jal, Sharif, Wang, Yang, Shabani}. It can also answer questions on cosmological tensions \cite{Sahoo1} and provide other astrophysical applications, such as  black holes \cite{Lavinia, Gogoi, Junior, Chen}, wormholes \cite{Sahoo2, Chakraborty, Rastgoo}, and   large scale structure \cite{Sahoo3, Nashed}. 

Starting from these considerations, the idea developed in this paper is to use the Noether theorem \cite{Bajardi}, applied only to the gravitational part of the Lagrangian action, which, from the action invariance   under global spacetime translations, will allow us to derive the associated Noether currents, which we could interpret as a possible expression of the gravitational energy-momentum pseudo-tensor in $f(Q)$ gravity. Hence, the local conservation  of gravitational field and matter contributions to the energy-momentum  will be analyzed on-shell.  In the low energy regime and in the coincident gauge, a  second-order approximation  of metric perturbations will be derived. Thus, a comparison will be discussed between pseudo-tensors in different theories such as $f(T)$ and $f(Q)$, fixing their respective gauges. Finally, it is shown that the Stokes theorem does not change for a  connection  in  general formulation with the external calculus of Cartan differential forms. In other words, the theorem is independent of the coordinates and the connection. For a useful comparison of various pseudo-tensor proposals and their conservation, see Refs.\cite{Goldberg:1958zz,Lee:1974nq,Multamaki:2007wb,Barraco:1998eq, Bak:1993us, Mikhail:1993kil,Lessner, Rosen:1994vj,Xulu:2002ix,Koivisto:2005yk, Iosifidis}. For a similar derivation of  gravitational pseudo-tensor in higher-order local, nonlocal, and metric teleparallel theories of gravity, see Refs.~\cite{CCT,CCT1,CCL,CCL1,ACCA,CCCL}. Furthermore, in order to calculate the power carried by gravitational waves and, in general, gravitational multipole formulas, it is necessary to study gravitational waves produced in various modified theories of gravitation, see Refs. ~\cite{CCN2,CC,CCL2,Capriolo,CCN1,CCC1,CCC2,Soudi:2018dhv,Hohmann:2018wxu}.

The paper is organized as follows. In Secs.~\ref{sec2} and \ref{sec3}, the basic notions of geometry related to non-metric compatible torsion theories are  provided, and then a synthetic analysis of  conserved quantities on manifolds, in particular, the Stokes theorem on boundary regions with non-metric connections with torsion are reported. In Sec.~\ref{sec4}, the energy-momentum pseudo-tensor for the gravitational field in modified symmetric teleparallel gravity $f(Q)$ is explicitly derived, and subsequently, in Sec.~\ref{sec5}, via a continuity equation, the local conservation of  energy and momentum densities on-shell is verified for  gravitational and non-gravitational fields. In Sec.~\ref{sec6}, a comparison is reported  between the gravitational pseudo-tensors of  metric teleparallel gravity $f(T)$ and the symmetric teleparallel gravity $f(Q)$, in Weitzenb\"ock  and coincident gauges, respectively.  In Sec.~\ref{sec7} we calculate the gravitational energy density for a Schwarzschild external solution in coincident gauge in STEGR gravity, while in Sec.~\ref{sec8}, we develop  the weak energy limit of  gravitational sector for the affine energy-stress tensor. The calculation is truncated at second order in the metric perturbations. Discussion and  conclusions are drawn in Sec.~\ref{sec9}.

\section{The geometric background}\label{sec2}
Spacetime can be modeled   as a four-dimensional Hausdorff and paracompact $C^{\infty}$ manifold $\mathcal{M}$ endowed with a non-degenerate pseudo-Riemannian metric tensor $g$ and a generic affine connection $\Gamma$, that is, the triplet $\{\mathcal{M},g,\Gamma\}$  defines the metric-affine geometry of the Universe. The two  objects $g$ and $\Gamma$ are, in general, independent. The most general affine connection $\Gamma$ is {\it torsioned} and {\it curved}, and we can express its torsion and curvature only in terms of  connection without  any metric. 

Let us first consider the covariant derivative with respect to  a general affine connection $\Gamma$ of a vector field $A^{\alpha}$ and a dual vector field $\omega_{\alpha}$ in the usual way, respectively as~\cite{BeltranJimenez:2018vdo}
\begin{equation}\label{0}
    \nabla_{\mu}A^{\alpha}=\partial_{\mu}A^{\alpha}+\Gamma^{\alpha}_{\phantom{\alpha}\mu\lambda}A^{\lambda}\ ,
\end{equation}
and 
\begin{equation}\label{0_1}
    \nabla_{\mu}\omega_{\alpha}=\partial_{\mu}\omega_{\alpha}-\Gamma^{\lambda}_{\phantom{\lambda}\mu\alpha}\omega_{\lambda}\ .
\end{equation}
Then, in coordinate basis, the torsion tensor $T^{\alpha}_{\phantom{\alpha}\mu\nu}$ is due to  the commutator of two covariant derivatives on the function $f$ as 
\begin{equation}\label{0_2}
    [\nabla_{\mu},\nabla_{\nu}]f=-T^{\lambda}_{\phantom{\lambda}\mu\nu}\nabla_{\lambda}f\ ,
\end{equation}
namely a tensor of type $(1,2)$, twice the antisymmetric part of the connection, 
\begin{equation}\label{1}
    T^{\alpha}_{\phantom{\alpha}\mu\nu}=2\Gamma^{\alpha}_{\phantom{\alpha}[\mu\nu]}\ ,
\end{equation}
with trace
\begin{equation}\label{1_1}
    T_{\mu}=T^{\alpha}_{\phantom{\alpha}\mu\alpha}\ .
\end{equation}

On the other hand, the Riemann tensor $R^{\alpha}_{\phantom{\alpha}\beta\mu\nu}$, always in the coordinate basis, is defined in terms of  the commutator $[\nabla_{\mu},\nabla_{\nu}]$ on the vector field $A^{\alpha}$, i.e., 
\begin{equation}\label{1_2}
    [\nabla_{\mu},\nabla_{\nu}]A^{\alpha}=R^{\alpha}_{\phantom{\alpha}\sigma\mu\nu}A^{\sigma}-T^{\lambda}_{\phantom{\lambda}\mu\nu}\nabla_{\lambda}A^{\alpha}\ ,
\end{equation}
that is, a tensor of type $(1,3)$, antisymmetric in its last two indices identified as 
\begin{equation}\label{2}
    R^{\alpha}_{\phantom{\alpha}\beta\mu\nu}=\partial_{\mu}\Gamma^{\alpha}_{\phantom{\alpha}\nu\beta}-\partial_{\nu}\Gamma^{\alpha}_{\phantom{\alpha}\mu\beta}+\Gamma^{\alpha}_{\phantom{\alpha}\mu\lambda}\Gamma^{\lambda}_{\phantom{\lambda}\nu\beta}-\Gamma^{\alpha}_{\phantom{\alpha}\nu\lambda}\Gamma^{\lambda}_{\phantom{\lambda}\mu\beta}\ ,
\end{equation}
\begin{equation}\label{3}
    R^{\alpha}_{\phantom{\alpha}\beta\mu\nu}=R^{\alpha}_{\phantom{\alpha}\beta[\mu\nu]}\ .
\end{equation}
The Ricci tensor is then obtained by contracting the first and the third index of the Riemann tensor as
\begin{equation}\label{4}
    R_{\mu\nu}=R^{\lambda}_{\phantom{\lambda}\mu\lambda\nu}\ .
\end{equation}
Finally, we have the Ricci scalar $R$ as the trace of the Ricci tensor $R_{\mu\nu}$.

In addition, since the connection is independent of the metric, it is non-metric, and the degree of non-compatibility of the affine connection with the metric is measured by the non-metricity tensor $Q_{\alpha\mu\nu}$ expressed as 
\begin{equation} \label{5}
    Q_{\alpha\mu\nu}=\nabla_{\alpha}g_{\mu\nu}=\partial_{\alpha}g_{\mu\nu}-\Gamma^{\lambda}_{\phantom{\lambda}\alpha\mu}g_{\lambda\nu}-\Gamma^{\lambda}_{\phantom{\lambda}\alpha\nu}g_{\lambda\mu}\ ,
\end{equation}
symmetric in its last two indices, 
\begin{equation}\label{6}
    Q_{\alpha\mu\nu}=Q_{\alpha(\mu\nu)}\ ,
\end{equation}
that can be contracted in two ways with $g_{\mu\nu}$ as
\begin{equation}\label{7}
    Q_{\alpha}=Q_{\alpha\phantom{\lambda}\lambda}^{\phantom{\alpha}\lambda}\ ,
\end{equation}
and 
\begin{equation}\label{8}
    \tilde{Q}_{\alpha}=Q^{\lambda}_{\phantom{\lambda}\lambda\alpha}\ ,
\end{equation}
where $\nabla_{\alpha}$ is the ordinary covariant derivative expressed with respect to the general affine connection $\Gamma$ in a holonomic basis of coordinates. Then, from Eqs.~\eqref{1},\eqref{2} and \eqref{5}, with a straightforward calculation, we get that a general affine connection can be uniquely decomposed in terms of the Christoffel symbols $\Bigl\{^{\,\alpha}_{\mu\nu}\Bigr\}$, the contorsion tensor $K^{\alpha}_{\phantom{\alpha}\mu\nu}$ and the disformation tensor $L^{\alpha}_{\phantom{\alpha}\mu\nu}$ as~\cite{Xu:2019sbp, Capozziello:2022zzh}
\begin{equation}\label{9}
    \Gamma^{\alpha}_{\phantom{\alpha}\mu\nu}=\Bigl\{^{\,\alpha}_{\mu\nu}\Bigr\}+K^{\alpha}_{\phantom{\alpha}\mu\nu}+L^{\alpha}_{\phantom{\alpha}\mu\nu}\ .
\end{equation}
The Christoffel symbols  $\Bigl\{^{\,\alpha}_{\mu\nu}\Bigr\}$ are defined as
\begin{equation}\label{10}
    \Bigl\{^{\,\alpha}_{\mu\nu}\Bigr\}=\frac{1}{2}g^{\alpha\beta}\left(\partial_{\mu}g_{\nu\beta}+\partial_{\nu}g_{\mu\beta}-\partial_{\beta}g_{\mu\nu}\right)\ .
\end{equation}
The contorsion tensor $K^{\alpha}_{\phantom{\alpha}\mu\nu}$ is defined as
\begin{equation}\label{11}
    K^{\alpha}_{\phantom{\alpha}\mu\nu}=\frac{1}{2}\;T^{\alpha}_{\phantom{\alpha}\mu\nu}+T^{\phantom{(\mu}\alpha}_{(\mu\phantom{\alpha}\nu)}\ ,
\end{equation}
antisymmetric in the first and third index, i.e., 
\begin{equation}\label{12}
    K_{\beta\mu\nu}=-K_{\mu\nu\beta}\ ,
\end{equation}
where the antisymmetric and symmetric parts of its last two indices are, respectively
\begin{equation}\label{13}
    K^{\alpha}_{\phantom{\alpha}[\mu\nu]}=\frac{1}{2} T^{\alpha}_{\phantom{\alpha}\mu\nu}=\Gamma^{\alpha}_{\phantom{\alpha}[\mu\nu]}\ ,\quad\text{and}\quad K^{\alpha}_{\phantom{\alpha}(\mu\nu)}= T^{\phantom{(\mu}\alpha}_{(\mu\phantom{\alpha}\nu)}\ .
\end{equation}
Therefore, its only possible contraction is
\begin{equation}\label{14}
   K^{\lambda}_{\phantom{\lambda}\lambda\mu}=-T_{\mu}\ .
\end{equation}
Finally, the disformation tensor $L^{\alpha}_{\phantom{\alpha}\mu\nu}$ is defined as
\begin{equation}\label{15}
    L^{\alpha}_{\phantom{\alpha}\mu\nu}=\frac{1}{2}\;Q^{\alpha}_{\phantom{\alpha}\mu\nu}-Q^{\phantom{(\mu}\alpha}_{(\mu\phantom{\alpha}\nu)}\ ,
\end{equation}
 symmetric in its two last indices, i.e.,
\begin{equation}\label{16}
     L^{\alpha}_{\phantom{\alpha}\mu\nu}= L^{\alpha}_{\phantom{\alpha}(\mu\nu)}\ ,
\end{equation}
with two possible contractions, 
\begin{equation}\label{17}
     L^{\lambda}_{\phantom{\lambda}\lambda\mu}=-\frac{1}{2}Q_{\mu}\quad\text{and}\quad L_{\mu\phantom{\lambda}\lambda}^{\phantom{\mu}\lambda}=\frac{1}{2}Q_{\mu}-\tilde{Q}_{\mu}\ .
\end{equation}
The only connection on pseudo-Riemannian manifold, symmetric and metric compatible, is the Levi-Civita (LC) one, i.e., from Eq.~\eqref{9},
\begin{equation}\label{18}
    \hat{\Gamma}^{\alpha}_{\phantom{\alpha}\mu\nu}=\Bigl\{^{\,\alpha}_{\mu\nu}\Bigr\}= \frac{1}{2}g^{\alpha\beta}\left(\partial_{\mu}g_{\nu\beta}+\partial_{\nu}g_{\mu\beta}-\partial_{\beta}g_{\mu\nu}\right)\,,
\end{equation}
where $\hat{\Gamma}^{\alpha}_{\phantom{\alpha}\mu\nu}$ is the symmetric-metric connection.

The non-metricity scalar $Q$ gives the  dynamical equivalence to GR. It is chosen as
\begin{equation}\label{19}
    Q=-\frac{1}{4}Q_{\alpha\mu\nu}Q^{\alpha\mu\nu}+\frac{1}{2}Q_{\alpha\mu\nu}Q^{\mu\alpha\nu}+\frac{1}{4}Q_{\alpha}Q^{\alpha}-\frac{1}{2}Q_{\alpha}\tilde{Q}^{\alpha}\ .
\end{equation}
Then,  the non-metricity conjugate tensor $P^{\alpha}_{\phantom{\alpha}\mu\nu}$,  i.e. the superpotential, is defined as 
\begin{equation}\label{19_1}
    P^{\alpha}_{\phantom{\alpha}\mu\nu}=\frac{1}{2}\frac{\partial Q}{\partial Q_{\alpha}^{\phantom{\alpha}\mu\nu}}=-\frac{1}{2}L^{\alpha}_{\phantom{\alpha}\mu\nu}+\frac{1}{4}g_{\mu\nu}\left(Q^{\alpha}-\tilde{Q}^{\alpha}\right)-\frac{1}{4}\delta^{\alpha}_{\phantom{\alpha}(\mu}Q_{\nu)}\ ,
\end{equation}
where 
\begin{equation}\label{19_1_1}
    P^{\alpha}_{\phantom{\alpha}\mu\nu}=P^{\alpha}_{\phantom{\alpha}(\mu\nu)}\ ,
\end{equation}
which allows us to write the non-metricity scalar $Q$ as
\begin{equation}\label{19_2}
Q=P^{\alpha}_{\phantom{\alpha}\mu\nu}Q_{\alpha}^{\phantom{\alpha}\mu\nu}\ .
\end{equation}
Similarly, we can introduce a scalar torsion $T$ as 
\begin{equation}\label{19_3}
    T=\biggl[\frac{1}{4}T_{\alpha\mu\nu}+\frac{1}{2}T_{\mu\alpha\nu}-g_{\alpha\mu}T_{\nu}\biggr]T^{\alpha\mu\nu}\ ,
\end{equation}
and thanks to the torsion conjugate tensor $S_{\alpha}^{\phantom{\alpha}\mu\nu}$ defined as 
\begin{equation}\label{19_3_1}
    S_{\alpha}^{\phantom{\alpha}\mu\nu}=\frac{1}{2}\frac{\partial T}{\partial T^{\alpha}_{\phantom{\alpha}\mu\nu}}=\frac{1}{4}T_{\alpha}^{\phantom{\alpha}\mu\nu}-\frac{1}{2}T^{[\mu\nu]}_{\phantom{\mu\nu}\alpha}-\delta_{\alpha}^{[\mu}T^{\nu]}\ ,
\end{equation}
we can write the scalar torsion as
\begin{equation}\label{19_4}
    T=S_{\alpha}^{\phantom{\alpha}\mu\nu}T^{\alpha}_{\phantom{\alpha}\mu\nu}\ .
\end{equation}
Then, it is possible to express the curvature $R$ of  connection $\Gamma$ and the curvature $\mathcal{R}$ of  Levi-Civita connection $\hat{\Gamma}$, in terms of  torsion scalar $T$,  non-metricity scalar $Q$ and the boundary term $B$ as
\begin{equation}\label{19_5}
    R=\mathcal{R}+T-Q-T_{\alpha\nu\mu}Q^{\mu\nu\alpha}-T_{\alpha}\bigl(Q^{\lambda}-\Tilde{Q}^{\lambda}\bigr)+\mathcal{D}_{\lambda}\bigl(Q^{\lambda}-\tilde{Q}^{\lambda}+2T^{\lambda}\bigr)\ ,
\end{equation}
where $\mathcal{D}$ is the covariant derivative expressed with respect to the Levi-Civita connection $\hat{\Gamma}$ and 
the boundary term is $B=\mathcal{D}_{\lambda}\bigl(Q^{\lambda}-\tilde{Q}^{\lambda}+2T^{\lambda}\bigr)$.
 
In a spacetime with a metric-flat, and torsioned connection, i.e. the  metric teleparallel gravity case,  under the assumptions 
\begin{equation}\label{19_6}
    Q_{\alpha\mu\nu}=0\ ,\quad\text{and}\quad R[\Gamma]=0\ ,
\end{equation}
 Eq.~\eqref{19_5} yields 
\begin{equation}\label{19_6_1}
    \mathcal{R}=-T-B\ ,
\end{equation}
with the boundary term 
\begin{equation}\label{19_6_2}
    B=2\mathcal{D}_{\lambda}T^{\lambda}\ .
\end{equation}
On the other hand, if the spacetime has a torsion-free and flat non-metric connection, we have the {\em symmetric teleparallel gravity} (STG). This is realized under the assumptions~\cite{BeltranJimenez:2019esp,Capozziello:2025hyw,Harko:2018gxr,BeltranJimenez:2019tme,Runkla:2018xrv,Nester:1998mp} 
\begin{equation}\label{19_6_3}
\boxed{
    T^{\alpha}_{\phantom{\alpha}\mu\nu}[\Gamma]=0\ ,\quad\text{and}\quad R^{\alpha}_{\phantom{\alpha}\beta\mu\nu}[\Gamma]=0\ .
    }
\end{equation}
In this case,  Eq.~\eqref{19_5} reads as~\cite{CCL2,De:2023xua,Runkla:2018xrv}
\begin{equation}\label{19_7}
    \mathcal{R}=Q-B\ ,
\end{equation}
and the boundary term  becomes
\begin{equation}\label{19_8}
    B=\mathcal{D}_{\alpha}\left(Q^{\alpha}-\tilde{Q}^{\alpha}\right)\ .
\end{equation}
From  assumptions~\eqref{19_6_3} on curvature and torsion,  the affine STG connection  can be written as
\begin{equation}\label{19_9}
    \Gamma^{\alpha}_{\phantom{\alpha}\mu\nu}=\frac{\partial x^{\alpha}}{\partial \xi^{\lambda}}\frac{\partial^2 \xi^{\lambda}}{\partial x^{\mu}\partial x^{\nu}}\ ,
\end{equation}
where $\xi^{\mu}=\xi^{\mu}(x^{\nu})$ is a collection of four functions in terms of a chart $\{x^{\nu}\}$. We can choose $\xi^{\mu}=x^{\mu}$ so that the connection $\Gamma^{\alpha}_{\phantom{\alpha}\mu\nu}$~\eqref{19_9} vanishes anywhere. It is worth noticing that there is still a residual gauge due to the freedom of choice of the $\xi^\mu$ because they are defined up to an affine transformation $x^{\mu}\rightarrow ax^{\mu}+b$ with $a$ and $b$ constants.  This fact allows us  to fix  the so-called {\em coincident gauge}  up to an affine transformation, where
\begin{equation}
    \Gamma^{\alpha}_{\phantom{\alpha}\mu\nu}=0\ ,
\end{equation}
everywhere on the manifold.  So in general, in symmetric teleparallel theories in the coincident gauge, some physical quantities, even if they are invariant under diffeomorphisms, can depend on the choice of  $\xi^{\mu}$, as we will see in STEGR. With these considerations in mind, we can consider  conserved quantities on  manifolds.
\section{Conserved quantities on manifolds}\label{sec3}
 We can formulate the Stokes theorem on the 4-dimensional submanifold $\Omega$ of  manifold $\mathcal M$ with a 3-dimensional boundary $\partial\Omega$, endowed with a generic curved, torsioned, and non-metric affine connection $\Gamma$. See Refs. ~\cite{Heisenberg:2023lru,Maggiore1}. In presence of non-metricity and torsion the divergence of the vector field $A^{\mu}$ is given by
 \begin{equation}\label{19_99}
     \nabla_{\mu}A^{\mu}=\mathcal{D}_{\mu}A^{\mu}+L_{\mu}A^{\mu}+K_{\mu}A^{\mu}\ ,
 \end{equation}
 with $\nabla$ covariant derivative with respect to non-metric and torsioned connection and $\mathcal{D}$ covariant derivative with respect to Levi Civita connection. 
 From Eqs.~\eqref{14} and \eqref{17}, taking into account Gauss theorem 
 \begin{equation}\label{19_99_1}
     \int_{\Omega}d^4x\; \partial_{\mu}\Bigl(\sqrt{-g}A^{\mu}\Bigr)=\int_{\partial\Omega}dS\, n_{\alpha}A^{\alpha}
     \ ,
 \end{equation}
 and the well-known relation
 \begin{equation}
     \partial_{\mu}\left(\sqrt{-g}A^{\mu}\right)=\sqrt{-g}\,\mathcal{D}_{\mu}A^{\mu}\ ,
 \end{equation}
and from Eq.~\eqref{19_99} we have
 \begin{equation}\label{19_100}
 \boxed{
     \int_{\Omega}d^4x\;\sqrt{-g}\biggl[\nabla_{\alpha}A^{\alpha}+\biggl(\frac{1}{2}Q_{\alpha}+T_{\alpha}\biggr)A^{\alpha}\biggr]=\int_{\partial\Omega}dS\, n_{\alpha}A^{\alpha}
     }\ .
 \end{equation}
 Here $A^{\alpha}$ is a vector field, $n_{\alpha}$ is a normal vector of the hypersurface $\partial\Omega$ and $dS$ is the invariant induced volume element on it. Now, we decompose the boundary $\partial\Omega$ with two constant time spacelike hypersurfaces $\Sigma_{1}$ and $\Sigma_{2}$ and a timelike hypersurface connecting them to the spatial infinity $\Sigma_{\infty}$, that is,
 \begin{equation}\label{19_101}
     \partial\Omega=\Sigma_{1}\cup\Sigma_{2}\cup\Sigma_{\infty}\ .
 \end{equation}
 Assuming that asymptotically in space the vector field $A^{\alpha}$ vanishes, the surface integral on $\Sigma_{\infty}$ also vanishes; hence, from Eq.\eqref{19_100}, we have
 \begin{equation}\label{19_102}
      \int_{\Omega}d^4x\;\sqrt{-g}\biggl[\nabla_{\alpha}A^{\alpha}+\biggl(\frac{1}{2}Q_{\alpha}+T_{\alpha}\biggr)A^{\alpha}\biggr]=\int_{\Sigma_{1}}dS\, n_{\alpha}A^{\alpha}-\int_{\Sigma_{2}}dS\, n_{\alpha}A^{\alpha}\ .
 \end{equation}
 The minus sign in the second surface integral in Eq.\eqref{19_102} is due to the orientation on $\Sigma_{2}$ inherited from $\Omega$, where its timelike normal vector $n_{\alpha}$ points inward, while  the timelike normal vector of hypersurface  $\Sigma_{1}$ points outward, that we assume to be positive. Then, contracting $\Gamma^{\alpha}_{\phantom{\alpha}\mu\nu}$ in Eq.~\eqref{9} in its first two indices with the metric $g_{\mu\nu}$, from Eqs.~\eqref{14} and the first of \eqref{17}, we get
 \begin{equation}\label{19_104}
     \Gamma_{\mu}=\frac{1}{\sqrt{-g}}\partial_{\mu}\sqrt{-g}-\frac{1}{2}Q_{\mu}-T_{\mu}\ ,
 \end{equation}
 while in absence of non-metricity and torsion, i.e., with $Q_{\mu}=T_{\mu}=0$, from Eq.~\eqref{19_104} we obtain the well-known contracted Levi-Civita connection 
 \begin{equation}\label{19_105}
     \hat{\Gamma}_{\mu}=\frac{1}{\sqrt{-g}}\partial_{\mu}\sqrt{-g}\ .
 \end{equation}
So, the covariant derivative of the scalar density of weight 1, $\sqrt{-g}$, from the following relation
 \begin{equation}\label{19_106}
     \nabla_{\mu}\sqrt{-g}=\partial_{\mu}\sqrt{-g}-\Gamma_{\mu}\sqrt{-g}\ ,
 \end{equation}
 yields
 \begin{equation}\label{19_107}
     \nabla_{\mu}\sqrt{-g}=\biggl(\frac{1}{2}Q_{\mu}+T_{\mu}\biggr)\sqrt{-g}\ .
 \end{equation}
 If the following balance equation for a vector field $A^{\alpha}$ in any connection is fulfilled
 \begin{equation}\label{19_103}
     \nabla_{\alpha}A^{\alpha}+\biggl(\frac{1}{2}Q_{\alpha}+T_{\alpha}\biggr)A^{\alpha}=0\ ,
 \end{equation}
 or equivalently if the integrand on the left of Eq.~\eqref{19_100} vanishes, from Eq.~\eqref{19_107} we obtain the vanishing of vector density  covariant divergence  $\sqrt{-g}A^{\mu}$, i.e., 
\begin{equation}\label{19_108}
    \nabla_{\mu}\Bigl(\sqrt{-g}A^{\mu}\Bigr)=0\ .
\end{equation}
From  Eq.\eqref{19_102}, we have the following conserved quantity $\mathcal{K}$ on four-dimensional submanifold $\Omega$, 
 \begin{equation}\label{19_110}
     \mathcal{K}=\int_{\Sigma} d^3y \sqrt{\vert\gamma\vert}\, n_{\alpha}A^{\alpha}\ ,
 \end{equation}
 with $\gamma$ the determinant of the induced metric $\gamma_{\mu\nu}$ on the 3-dimensional hypersurface $\Sigma$ in $y_{\mu}$ coordinates. Finally, from Eqs.~\eqref{19_100} and \eqref{19_107}, the Stokes theorem on a four-dimensional submanifold $\Omega$ with a three-dimensional boundary $\partial\Omega$, endowed with a generic affine connection $\Gamma$, can be rewritten as 
 \begin{equation}\label{19_111}
 \boxed{
      \int_{\Omega}d^4x\; \nabla_{\mu}\Bigl(\sqrt{-g}A^{\mu}\Bigr)=\int_{\partial\Omega} d^3y \sqrt{\vert\gamma\vert}\, n_{\alpha}A^{\alpha}
      }\ .
 \end{equation}
Eq.~\eqref{19_111} verifies a standard result in differential geometry, where the Stokes theorem in its intrinsic coordinate-free formulation can be expressed in terms of $p$-forms with the Cartan exterior calculus, and is metric- and connection-independent.

\section{The gravitational energy-momentum pseudo-tensor for $f(Q)$ non-metric gravity}\label{sec4}
Let us now take into account   a  non-linear extension of STEGR, via a generic analytic function $f$ of the non-metricity scalar $Q$. The  action of  $f(Q)$  gravity, including a matter term, can be expressed as follows 
\begin{equation}\label{20}
S=\int_{\Omega}d^{4}x\,\Bigl[\frac{1}{2\kappa^{2}}\sqrt{-g}f\left(Q\right)+\lambda_{\alpha}^{\phantom{\alpha}\beta\mu\nu}R^{\alpha}_{\phantom{\alpha}\beta\mu\nu}+\lambda_{\alpha}^{\phantom{\alpha}\mu\nu}T^{\alpha}_{\phantom{\alpha}\mu\nu}+\sqrt{-g}\mathcal{L}_{m}\bigl(g\bigr)\Bigr]\ ,
\end{equation}
where $\kappa^{2}=8\pi G/c^{4}$ and the Lagrange multipliers $\lambda_{\alpha}^{\phantom{\alpha}\beta\mu\nu}=\lambda_{\alpha}^{\phantom{\alpha}\beta[\mu\nu]}$, $\lambda_{\alpha}^{\phantom{\alpha}\mu\nu}=\lambda_{\alpha}^{\phantom{\alpha}[\mu\nu]}$ are tensor densities of weight $w=+1$, antisymmetric in their last two indices. They are necessary to impose the two  curvature-free and torsion-free conditions where further 96 plus 24 independent scalar fields have to be considered.  To obtain the field equations, we have to vary  action~\eqref{20}, assuming a null variation of  fields on the boundary, and then, from $\delta_{g} S=0$, varying with respect to the metric $g_{\mu\nu}$.  The metric equations are~\cite{CCN2,CC} 
\begin{equation}\label{20_1}
\frac{2}{\sqrt{-g}}\nabla_{\alpha}\left(\sqrt{-g}f_{Q}P^{\alpha}_{\phantom{\alpha}\mu\nu}\right)-\frac{1}{2}g_{\mu\nu}f+f_{Q}\Bigl(P_{\mu\alpha\beta}Q_{\nu}^{\phantom{\nu}\alpha\beta}-2Q^{\alpha\beta}_{\phantom{\alpha\beta}\mu}P_{\alpha\beta\nu}\Bigr)=\kappa^{2}T_{\mu\nu}\ ,
\end{equation}
where $f_{Q}=\partial f/\partial Q$, $\nabla$ is the covariant derivative with respect to the symmetric teleparallel connection $\Gamma$. The matter energy-momentum tensor is defined as 
\begin{equation}\label{20_2}
T_{\mu\nu}=-\frac{2}{\sqrt{-g}}\frac{\delta S_{m} }{\delta g^{\mu\nu}}\ ,
\end{equation}
where $S_{m}$ is the material Lagrangian, that is 
\begin{equation}
S_{m}=\int_{\Omega}d^{4}x\,\sqrt{-g}\mathcal{L}_{m}\ .
\end{equation}
By the stationary condition $\delta_{\Gamma} S=0$, varying Eq.~\eqref{20} with respect to STG connection,  we find the following connection equations 
\begin{equation}\label{20_3}
\nabla_{\mu}\nabla_{\nu}\Bigl[\sqrt{-g}f_{Q}P^{\mu\nu}_{\phantom{\mu\nu}\alpha}\Bigr]=0\ .
\end{equation}
Finally, by varying the action~\eqref{20} with respect to the Lagrange multipliers $\lambda_{\alpha}^{\phantom{\alpha}\beta\mu\nu}$ and $\lambda_{\alpha}^{\phantom{\alpha}\mu\nu}$, imposing stationary conditions $\delta_{\lambda_{\alpha}^{\phantom{\alpha}\beta\mu\nu}} S=0$ and $\delta_{\lambda_{\alpha}^{\phantom{\alpha}\mu\nu}} S=0$,  constraints in STG are derived, that is 
\begin{equation}\label{20_4}
R^{\alpha}_{\phantom{\alpha}\beta\mu\nu}=0\ ,
\end{equation}
\begin{equation}\label{20_5}
T^{\alpha}_{\phantom{\alpha}\mu\nu}=0\ .
\end{equation}
To obtain the gravitational energy and momentum, let us vary only the free gravitational action $S_{g}$ with respect to the generic coordinates $x^{\mu}$ without imposing any conditions on the boundary of the integration domain $\Omega$, which we indicate as $\tilde{\delta}$. The variation with fixed coordinates is indicated with $\delta$, i.e., 
\begin{equation}\label{21}
\tilde{\delta}S_{g}=\int_{\Omega}d^{4}x\,\Biggl[\frac{1}{2\kappa^{2}}\delta\bigl(\sqrt{-g}f\left(Q\right)\bigr)+\delta\bigl(\lambda_{\alpha}^{\phantom{\alpha}\beta\mu\nu}R^{\alpha}_{\phantom{\alpha}\beta\mu\nu}\bigr)+\delta\bigl(\lambda_{\alpha}^{\phantom{\alpha}\mu\nu}T^{\alpha}_{\phantom{\alpha}\mu\nu}\bigr)+\partial_{\lambda}\bigl(\sqrt{-g}\mathcal{L}_{g}\delta x^{\lambda}\bigr)\Biggr]\ .
\end{equation}
Taking into account the following variations
\begin{align}
\delta \sqrt{-g}&=-\frac{1}{2}\sqrt{-g}g_{\mu\nu}\delta g^{\mu\nu}\label{22}\ ,\\
\delta Q&=\Bigl(P_{\mu\alpha\beta}Q_{\nu}^{\phantom{\nu}\alpha\beta}-2Q^{\alpha\beta}_{\phantom{\alpha\beta}\mu}P_{\alpha\beta\nu}\Bigr)\delta g^{\mu\nu}-2P^{\alpha}_{\phantom{\alpha}\mu\nu}\nabla_{\alpha}\delta g^{\mu\nu}-4P^{\mu\nu}_{\phantom{\mu\nu}\alpha}\delta\Gamma^{\alpha}_{\phantom{\alpha}\mu\nu}\label{23}\ ,\\
\delta R^{\alpha}_{\phantom{\alpha}\beta\mu\nu}&=2\nabla_{[\mu}\delta\Gamma^{\alpha}_{\phantom{\alpha}\nu]\beta}+T^{\gamma}_{\phantom{\gamma}\mu\nu}\delta\Gamma^{\alpha}_{\phantom{\alpha}\gamma\beta}\label{24}\ ,\\
\delta T^{\alpha}_{\phantom{\alpha}\mu\nu}&=2\delta\Gamma^{\alpha}_{\phantom{\alpha}[\mu\nu]}\ ,\label{25}
\end{align}
and the Leibniz rule for derivatives, the local variation~\eqref{21} becomes 
\begin{align}\label{26}
\tilde{\delta}S_{g}=&\int_{\Omega}d^{4}x\,\frac{\delta S_{g}}{\delta g^{\mu\nu}}\delta g^{\mu\nu}+\frac{\delta S_{g}}{\delta \Gamma^{\alpha}_{\phantom{\alpha}\mu\nu}}\delta \Gamma^{\alpha}_{\phantom{\alpha}\mu\nu}+\frac{\delta S_{g}}{\delta\lambda_{\alpha}^{\phantom{\alpha}\beta\mu\nu}}\delta\lambda_{\alpha}^{\phantom{\alpha}\beta\mu\nu}+\frac{\delta S_{g}}{\delta\lambda_{\alpha}^{\phantom{\alpha}\mu\nu}}\delta\lambda_{\alpha}^{\phantom{\alpha}\mu\nu}\nonumber\\
&-\frac{1}{2\kappa^{2}}\nabla_{\alpha}\Bigl(\sqrt{-g}f_{Q}P^{\alpha}_{\phantom{\alpha}\mu\nu}\delta g^{\mu\nu}\Bigr)+\nabla_{\mu}\Bigl(2\lambda_{\alpha}^{\phantom{\alpha}\beta\mu\nu}\delta\Gamma^{\alpha}_{\phantom{\alpha}\nu\beta}\Bigr)+\partial_{\lambda}\bigl(\sqrt{-g}\mathcal{L}_{g}\delta x^{\lambda}\bigr)\ ,
\end{align}
where the functional derivative with respect to  the metric tensor is
\begin{equation}
    \frac{\delta S_{g}}{\delta g^{\mu\nu}}=\frac{1}{2\kappa^2}\biggl[2\nabla_{\alpha}\left(\sqrt{-g}f_{Q}P^{\alpha}_{\phantom{\alpha}\mu\nu}\right)-\frac{1}{2}\sqrt{-g}g_{\mu\nu}f+\sqrt{-g}\,f_{Q}\Bigl(P_{\mu\alpha\beta}Q_{\nu}^{\phantom{\nu}\alpha\beta}-2Q^{\alpha\beta}_{\phantom{\alpha\beta}\mu}P_{\alpha\beta\nu}\Bigr)\biggr]\ ,
\end{equation}
and those with respect to the connection and Lagrange multipliers are, respectively, 
\begin{equation}
    \frac{\delta S_{g}}{\delta \Gamma^{\alpha}_{\phantom{\alpha}\mu\nu}}=2\biggl[\nabla_{\beta}\lambda_{\alpha}^{\phantom{\alpha}\nu\mu\beta}+\lambda_{\alpha}^{\phantom{\alpha}\mu\nu}-\frac{\sqrt{-g}}{\kappa^2}f_{Q}P^{\mu\nu}_{\phantom{\mu\nu}\alpha}\biggr]\ ,
\end{equation}
\begin{equation}
\frac{\delta S_{g}}{\delta\lambda_{\alpha}^{\phantom{\alpha}\beta\mu\nu}}=R^{\alpha}_{\phantom{\alpha}\beta\mu\nu}\ ,
\end{equation}
\begin{equation}
\frac{\delta S_{g}}{\delta\lambda_{\alpha}^{\phantom{\alpha}\mu\nu}}=T^{\alpha}_{\phantom{\alpha}\mu\nu}\ .
\end{equation}
According to the following integral identity 
\begin{equation}\label{27}
\int_{\Omega}\nabla_{\alpha}\bigl(\sqrt{-g}A^{\alpha}\bigr)d^{4}x=\int_{\Omega}d^{4}x\sqrt{-g}\mathcal{D}_{\alpha}A^{\alpha}
=\int_{\Omega}d^{4}x\,\partial_{\alpha}\bigl(\sqrt{-g}A^{\alpha}\bigr)\ ,
\end{equation}
where $\mathcal{D}_{\alpha}$ is the Levi-Civita covariant derivative, taking into account the metric field Eqs.~\eqref{20_1} in vacuum without matter because we are interested in deriving the energy-momentum of the gravitational field only, the  remaining field Eqs.~\eqref{20_3}--\eqref{20_5},  action~\eqref{26} takes the following form
\begin{equation}\label{28}
 \tilde{\delta}S_{g}=\int_{\Omega}d^{4}x\,\partial_{\alpha}\biggl(-\frac{1}{2\kappa^{2}}\sqrt{-g}f_{Q}P^{\alpha}_{\phantom{\alpha}\mu\nu}\delta g^{\mu\nu}+2\lambda_{\omega}^{\phantom{\omega}\beta\alpha\nu}\delta\Gamma^{\omega}_{\phantom{\alpha}\nu\beta}+\frac{\sqrt{-g}}{2\kappa^{2}}f(Q)\delta^{\alpha}_{\lambda}\delta x^{\lambda}\biggr)\ .
\end{equation}
In order to use the Noether theorem, let us consider infinitesimal global spacetime translations as
\begin{equation}\label{29}
x^{\prime\mu}=x^{\mu}+\epsilon^{\mu}\ .
\end{equation}
The inverse metric tensor $g^{\mu\nu}$ and the connection $\Gamma^{\alpha}_{\phantom{\alpha}\mu\nu}$ change as
\begin{equation}\label{29_1}
\delta g^{\mu\nu}=-g^{\mu\nu}_{\phantom{\mu\nu},\lambda}\epsilon^{\lambda}\ ,
\end{equation}
\begin{equation}
\delta\Gamma^{\alpha}_{\phantom{\alpha}\mu\nu}=-\Gamma^{\alpha}_{\phantom{\alpha}\mu\nu,\lambda}\epsilon^{\lambda}\ ,
\end{equation}
with 
\begin{equation}
\delta x^{\mu}=\epsilon^{\mu}\ .
\end{equation}
If  only the gravitational part of  action is invariant under global translations $\tilde{\delta}S_{g}=0$, we derive the locally conserved affine tensor, namely the following local continuity equation 
\begin{equation}
\partial_{\alpha}\bigl(\sqrt{-g}J^{\alpha}\bigr)=0\ ,
\end{equation}
with the Noether current $J^{\alpha}$ defined as
\begin{equation}
J^{\alpha}=\tau^{\alpha}_{\phantom{\alpha}\lambda|f(Q)}\epsilon^{\lambda}\ ,
\end{equation}
or, equivalently, for the independence of $\epsilon^{\lambda}$ from the coordinates 
\begin{equation}
\partial_{\alpha}\bigl(\sqrt{-g}\tau^{\alpha}_{\phantom{\alpha}\lambda|f(Q)}\bigr)=0\ .
\end{equation}
We can then identify the Noether current for a given $\epsilon^{\lambda}$ as the {\em gravitational energy-momentum pseudo-tensor of $f(Q)$ gravity}. Without gauge fixing, it became
\begin{equation}\label{30}
\boxed{
\tau^{\alpha}_{\phantom{\alpha}\lambda|f(Q)}=\frac{1}{2\kappa^{2}}\Bigl[f_{Q}P^{\alpha}_{\phantom{\alpha}\mu\nu}g^{\mu\nu}_{\phantom{\mu\nu},\lambda}+f(Q)\delta^{\alpha}_{\lambda}-\frac{4\kappa^{2}}{\sqrt{-g}}\lambda_{\omega}^{\phantom{\omega}\beta\alpha\nu}\,\Gamma^{\omega}_{\phantom{\omega}\nu\beta,\lambda}\Bigr]
}\ ,
\end{equation}
where the Lagrange multiplier $\lambda_{\omega}^{\phantom{\omega}\beta\alpha\nu}$ is related to the generalized momentum $\pi_{\omega}^{\phantom{\omega}\beta\alpha\nu}$ conjugated to the symmetric teleparallel connection $\Gamma^{\omega}_{\phantom{\omega}\nu\beta}$, defined as
\begin{equation}\label{31}
\pi_{\omega}^{\phantom{\omega}\beta\alpha\nu}=\frac{\partial \mathcal{L}_{g}}{\partial\bigl(\partial_{\alpha}\Gamma^{\omega}_{\phantom{\omega}\nu\beta}\bigr)}\ .
\end{equation}
In fact, we have
\begin{equation}\label{32}
\lambda_{\omega}^{\phantom{\omega}\beta\alpha\nu}=\frac{\partial L_{g}}{\partial R^{\omega}_{\phantom{\omega}\beta\alpha\nu}}=\frac{1}{2}\frac{\partial L_{g}}{\partial\bigl(\partial_{\alpha}\Gamma^{\omega}_{\phantom{\omega}\nu\beta}\bigr)}=\frac{\sqrt{-g}}{2}\pi_{\omega}^{\phantom{\omega}\beta\alpha\nu}\ \ ,
\end{equation}
with $L_{g}=\sqrt{-g}\mathcal{L}_{g}$.
Then, if we define the momentum conjugated to the metric tensor $g_{\mu\nu}$ as 
\begin{equation}\label{33}
\widetilde{\pi}^{\alpha}_{\phantom{\alpha}\mu\nu}=\frac{\partial \mathcal{L}_{g}}{\partial\bigl(\partial_{\alpha}g^{\mu\nu}\bigr)}\ ,
\end{equation}
we get
\begin{equation}\label{34}
\widetilde{\pi}^{\alpha}_{\phantom{\alpha}\mu\nu}=-\frac{f_{Q}}{\kappa^2}P^{\alpha}_{\phantom{\alpha}\mu\nu}\ .
\end{equation}
From  Eq.~\eqref{30}, the gravitational energy-momentum pseudo-tensor of $f(Q)$ gravity, where we have turn off both the curvature $R$ and the torsion $T$, can be rewritten as
\begin{equation}\label{35}
\boxed{
-\tau^{\alpha}_{\phantom{\alpha}\lambda|f(Q)}=\Biggl[\frac{\widetilde{\pi}^{\alpha}_{\phantom{\alpha}\mu\nu}}{2}\partial_{\lambda}g^{\mu\nu}+\pi_{\omega}^{\phantom{\omega}\beta\alpha\nu}\partial_{\lambda}\Gamma^{\omega}_{\phantom{\omega}\nu\beta}-\mathcal{L}_{g}\delta^{\alpha}_{\lambda}\Biggr]_{R=T=0}
}\ ,
\end{equation} 
that is, the time-time component of  Eq.~\eqref{35}  can be regarded as a sort of Hamiltonian associated with the gravitational part, which represents the {\em gravitational energy of f(Q) non-metric gravity in any gauge}, namely
\begin{equation}
-\tau^{0}_{\phantom{0}0|f(Q)}=\Biggl[\frac{\widetilde{\pi}^{0}_{\phantom{0}\mu\nu}}{2}\partial_{0}g^{\mu\nu}+\pi_{\omega}^{\phantom{\omega}\beta 0\nu}\partial_{0}\Gamma^{\omega}_{\phantom{\omega}\nu\beta}-\mathcal{L}_{g}\Biggr]_{R=T=0}\ ,
\end{equation}
apart from the one-half factor due to the definition of the non-metricity conjugate $P^{\alpha}_{\phantom{\alpha}\mu\nu}$~\eqref{19_1}.

It is worth noticing that using Noether's theorem for local displacement transformations, some covariant quantities have been derived in symmetric teleparallel theories such as STEGR  as in Refs.~\cite{Emtsova:2022uij,Emtsova:2025mdp}. Among these results, the gravitational energy-momentum tensor is noteworthy. It differs from the one obtained in the present paper because it is a covariant object, while the derived energy-momentum pseudo-tensor as well as in Ref.~\cite{Gomes:2022vrc}, is only an affine quantity, having been obtained for global displacement transformations.

\section{The conservation of  energy-momentum complex pseudo-tensor in $f(Q)$ non-metric gravity}\label{sec5}

 Let us show now that the total energy and momentum densities, including gravitational and matter contributions, are locally conserved.  To do this end, we consider the full Eqs.~\eqref{20_1} where, by raising the first index, we can rewrite them as 
\begin{equation}\label{40}
\mathcal{F}^{\mu}_{\phantom{\mu}\nu}=\kappa^2 T^{\mu}_{\phantom{\mu}\nu}\ ,
\end{equation}
where 
\begin{equation}\label{41}
\mathcal{F}^{\mu}_{\phantom{\mu}\nu}=\frac{2}{\sqrt{-g}}\nabla_{\alpha}\left(\sqrt{-g}f_{Q}P^{\alpha\mu}_{\phantom{\alpha\mu}\nu}\right)-\frac{1}{2}\delta^{\mu}_{\nu}f+f_{Q}P^{\mu}_{\phantom{\mu}\alpha\beta}Q_{\nu}^{\phantom{\nu}\alpha\beta}\ .
\end{equation}
Furthermore, from the explicit expression of gravitational pseudo-tensor~\eqref{30}, the local variation of  action~\eqref{26}, for an infinitesimal global translation, can be rewritten as 
\begin{equation}\label{42}
\tilde{\delta}S_{g}=\int_{\Omega}d^{4}x\,\biggl[\frac{\delta S_{g}}{\delta g^{\mu\nu}}\delta g^{\mu\nu}+\frac{\delta S_{g}}{\delta \Gamma^{\alpha}_{\phantom{\alpha}\mu\nu}}\delta \Gamma^{\alpha}_{\phantom{\alpha}\mu\nu}+\partial_{\alpha}\bigl(\sqrt{-g}\tau^{\alpha}_{\phantom{\alpha}\lambda|f(Q)}\bigr)\epsilon^{\lambda}\biggr]\ .
\end{equation}
From Eqs~\eqref{20_1}--\eqref{20_5}, the variation~\eqref{42} becomes 
\begin{equation}\label{43}
\tilde{\delta}S_{g}=\int_{\Omega}d^{4}x\,\biggl[\frac{\sqrt{-g}}{2}T_{\mu\nu}\delta g^{\mu\nu}+\partial_{\alpha}\bigl(\sqrt{-g}\tau^{\alpha}_{\phantom{\alpha}\lambda|f(Q)}\bigr)\epsilon^{\lambda}\biggr]\ .
\end{equation}
From the invariance under global translations of  gravitational action and the variation~\eqref{29_1}, we have, for any $\epsilon^{\lambda}$,
\begin{equation}\label{44}
\frac{1}{2}\sqrt{-g}T^{\mu\nu}g_{\mu\nu,\lambda}+\partial_{\alpha}\bigl(\sqrt{-g}\tau^{\alpha}_{\phantom{\alpha}\lambda|f(Q)}\bigr)=0\ .
\end{equation}
Considering the following relation
\begin{equation}\label{45}
\sqrt{-g}\mathcal{D}_{\alpha}T^{\alpha}_{\phantom{\alpha}\lambda}=\partial_{\alpha}\bigl(\sqrt{-g}T^{\alpha}_{\phantom{\alpha}\lambda}\bigr)-\frac{1}{2}\sqrt{-g}T^{\mu\nu}g_{\mu\nu,\lambda}\ ,
\end{equation}
Eq.~\eqref{44} assumes the form
\begin{equation}\label{46}
\partial_{\alpha}\Bigl[\sqrt{-g}\bigl(T^{\alpha}_{\phantom{\alpha}\lambda}+\tau^{\alpha}_{\phantom{\alpha}\lambda}\bigr)\Bigr]=\sqrt{-g}\mathcal{D}_{\alpha}T^{\alpha}_{\phantom{\alpha}\lambda}\ .
\end{equation}
Since, from the connection Eq. \eqref{20_2}, in $f(Q)$ non-metric gravity,  the LC covariant divergence of the left-hand side of Eq.~\eqref{40} vanishes, i.e.,~\cite{CCL2} 
\begin{equation}\label{47}
\mathcal{D}_{\alpha}\mathcal{F}^{\alpha}_{\phantom{\alpha}\lambda}=0\ ,
\end{equation}
the energy-momentum tensor $T^{\alpha}_{\phantom{\alpha}\lambda}$ is LC covariantly conserved on shell, i.e., $\mathcal{D}_{\alpha}T^{\alpha}_{\phantom{\alpha}\lambda}=0$. Then, also the energy-momentum complex pseudo-tensor $\mathcal{T}^{\alpha}_{\phantom{\alpha}\lambda}$ in $f(Q)$ non-metric gravity is locally conserved on-shell
\begin{equation}\label{48}
\boxed{
\partial_{\alpha}\mathcal{T}^{\alpha}_{\phantom{\alpha}\lambda}=\partial_{\alpha}\Bigl[\sqrt{-g}\bigl(T^{\alpha}_{\phantom{\alpha}\lambda}+\tau^{\alpha}_{\phantom{\alpha}\lambda}\bigr)\Bigr]=0
}\ ,
\end{equation}
where 
\begin{equation}\label{49}
\mathcal{T}^{\alpha}_{\phantom{\alpha}\lambda}=T^{\alpha}_{\phantom{\alpha}\lambda}+\tau^{\alpha}_{\phantom{\alpha}\lambda}\ .
\end{equation}
The energy and momentum of gravitational and matter fields of an isolated system, immersed in an asymptotically flat 3-dimensional spacelike region $\Sigma$ of a manifold $\mathcal M$, can be defined as
\begin{equation}\label{49_1}
    P^{\alpha}=\int_{\Sigma}d^3x\;\Bigl[\sqrt{-g}\bigl(T^{\alpha}_{\phantom{\alpha}0}+\tau^{\alpha}_{\phantom{\alpha}0}\bigr)\Bigr]\ ,
\end{equation}
where a spacetime is asymptotically flat or shows asymptotic flatness if there is any  coordinates system $\{x^{\mu}\}$ such that the metric components in these coordinates behave like $g_{\mu\nu}=\eta_{\mu\nu}+\mathcal{O}(1/r)$ as $r\rightarrow\infty$ along spatial or null directions, which for a three-dimensional spacelike hypersurface reduces to the following asymptotic behavior $g_{ij}=\delta_{ij}+\mathcal{O}(1/r)$. For more details see Ref.~\cite{Wald} and for a  rigorous and formal definition  see Ref.~\cite{ASH}. Note that, although the integrand of \eqref{49_1} is an affine tensor density, the integrated quantity $P^{\alpha}$ is a covariant vector, that is, an invariant gauge object because it is integrated over asymptotically flat regions where all of the fields vanish. That is, integration on asymptotically flat spaces eliminates the affine character of physical quantities, making them gauge invariant. 

\section{Comparing $f(Q)$ and $f(T)$ gravitational energy-momentum pseudo-tensors}\label{sec6}
The metric teleparallel  $f(T)$ gravity is  a non-linear extension of TEGR. It is formulated in terms of a tetrad basis $\{{\bf E}_{a}\}$, a local basis orthonormal with respect to the Minkowski metric, i.e., 
\begin{equation}\label{50}
{\bf E}_{a}=E^{\mu}_{\phantom{\mu}a}{\bf e}_{\mu}\ ,
\end{equation}
with 
\begin{equation}\label{51}
\eta_{ab}=E^{\mu}_{\phantom{\mu}a}E^{\nu}_{\phantom{\nu}b}g_{\mu\nu}\ ,
\end{equation}
or, in terms of its dual tetrad basis $\{\boldsymbol{\theta}^{a}\}$,
\begin{equation}
\boldsymbol{\theta}^{a}=E_{\mu}^{\phantom{\mu}a}\boldsymbol{\theta}^{\mu}\ ,
\end{equation}
with
\begin{equation}\label{52}
g_{\mu\nu}=E_{\mu}^{\phantom{\mu}a}E_{\nu}^{\phantom{\nu}b}\eta_{ab}\ ,
\end{equation}
 where $\{{\bf e}_{\mu}\}$ and $\{\boldsymbol{\theta}^{\mu}\}$ are the ordinary coordinate basis vectors and coordinate basis one-forms, respectively.  Here, we assume the tetrad postulate of  absolute parallelism, that is we choose a class of frames where the spin connection vanishes. This implies the following Weitzenb\"ock connection $^{(w)}\Gamma^{\alpha}_{\phantom{\alpha}\mu\nu}$
 \begin{equation}\label{53}
 ^{(w)}\Gamma^{\alpha}_{\phantom{\alpha}\mu\nu}=E^{\alpha}_{\phantom{\alpha}a}\partial_{\nu}E_{\mu}^{\phantom{\mu}a}\ ,
 \end{equation}
 that is, a flat metric-compatible connection  \cite{Pereira}. 
 Hence, the gravitational energy-momentum pseudo-tensor for $f(T)$ gravity  is~\cite{CCT1} 
\begin{equation}\label{54}
\tau^{\alpha}_{\phantom{\alpha}\lambda|f(T)}=-\frac{1}{2\kappa^{2}}\Bigl[4f_{T}S_{a}^{\phantom{a}\rho\alpha}E^{\phantom{\rho}a}_{\rho\phantom{a},\lambda}+f(T)\delta^{\alpha}_{\lambda}\Bigr]\ ,
\end{equation}
where the superpotential $S_{a}^{\phantom{a}\rho\alpha}$ is defined as
\begin{equation}\label{55}
S_{a}^{\phantom{a}\rho\alpha}=-\frac{1}{4e}\frac{\partial (e\,T)}{\partial\bigl(\partial_{\alpha} E^{a}_{\phantom{a}\rho}\bigr)}\ ,
\end{equation}
with $e=\text{det}(E_{\rho}^{\phantom{\rho}a})$.  

In the case of  symmetric teleparallel  $f(Q)$ gravity, the gravitational energy-momentum pseudo-tensor~\eqref{30}, in the coincident gauge, becomes
\begin{equation}\label{56}
\boxed{
\tau^{\alpha}_{\phantom{\alpha}\lambda|f(Q)}=\frac{1}{2\kappa^{2}}\Bigl[f_{Q}P^{\alpha}_{\phantom{\alpha}\mu\nu}g^{\mu\nu}_{\phantom{\mu\nu},\lambda}+f(Q)\delta^{\alpha}_{\lambda}\Bigr]
}\ .
\end{equation}
It is worth noticing that in Eq. \eqref{30}, the partial derivative of the connection vanishes in the coincident gauge, because, in this coordinate system, the connection vanishes not only at one point but in the whole region of the manifold, or, more generally, everywhere on the manifold.

In teleparallel gravity,   dynamical variables are the components of the tetrad field $E^{\mu}_{\phantom{\mu}a}$ while, in  non-metric symmetric teleparallel  gravity (in particular in the coincident gauge),  dynamical variables are the components of  metric tensor $g_{\mu\nu}$. More precisely, the $(0,2)$ tensor field, symmetric and non-degenerate, with $(1,3)$-signature, the Lorentz metric $\boldsymbol{g}$, has  components $g_{\mu\nu}$ and  inverse components $g^{\mu\nu}$
\begin{equation}\label{58}
{\bf g}=g_{\mu\nu}\boldsymbol{\theta}^{\mu}\otimes\boldsymbol{\theta}^{\nu}=g^{\mu\nu}{\bf e}_{\mu}\otimes{\bf e}_{\nu}\ ,
\end{equation}
where $\otimes$ is the ordinary tensor product. Therefore, the two pseudo-tensors of $f(T)$, in tetrad frame~\eqref{54}, and $f(Q)$, in coincident gauge~\eqref{56}, have the same structure. In fact, they both depend on their  conjugate tensors, the torsion conjugate $S_{a}^{\phantom{a}\rho\alpha}$ and the non-metricity conjugate $P^{\alpha}_{\phantom{\alpha}\mu\nu}$, respectively. 

Specifically, they depend, in the same way, on their respective inverse   dynamical variables, that is the components of the inverse tetrad field $E_{\mu}^{\phantom{\mu}a}$ and the components of the  inverse metric tensor $g^{\mu\nu}$. Furthermore, they depend, in the same way,  on their respective scalars, the torsion scalar $T$ and the non-metricity scalar $Q$ through the analytic functions $f(T)$ and $f(Q)$.  

\section{Gravitational energy for Schwarzschild external solution in coincident gauge in STEGR gravity}\label{sec7}

As an application of the gravitational energy-momentum pseudo-tensor in Eq.~\eqref{30}, we derive, in STEGR gravity where $f(Q)=Q$, for the spherically symmetric solution of the field equations in vacuum, i.e., the Schwarzschild solution, the gravitational energy in spherical coordinates in coincident gauge. Generally, in $f(Q)$ theories, the static and spherically symmetric solution in spherical coordinates is not compatible with the coincident gauge, because otherwise the connection does not satisfy the symmetry conditions~\cite{Bahamonde:2022zgj}. However, in $f(Q)=Q$. i.e. in STEGR, it is always possible to find a spherical coordinate system that trivializes the connection and the Schwarzschild solution, in these coordinates,  satisfies the field equations. Let us consider the following external Schwarzschild solution in spherical coordinates $(t,r,\theta,\varphi)=(x^0,x^1,x^2,x^3)$, that is the metric element
\begin{equation}\label{60.01}
ds^2=\bigl(1-\frac{r_{s}}{r}\bigr)c^2dt^2-\frac{1}{1-\frac{r_{s}}{r}}dr^2-r^2\bigl(d\theta^2+\sin^2\theta d\varphi^2\bigr)\ ,
\end{equation}
with $r>r_{g}$ where $r_{s}=2MG/c^2$ is the Schwarzschild radius. Furthermore, we rewrite  Eq.~\eqref{20_1} in vacuum as
\begin{equation}\label{60.01_1}
E_{\mu\nu}=f_{Q}\hat{G}_{\mu\nu}+\frac{1}{2}g_{\mu\nu}\left(f_{Q}Q-f\right)+2f_{QQ}\partial_{\lambda}QP^{\lambda}_{\phantom{\lambda}\mu\nu}=0\ ,
\end{equation}
with $\hat{G}_{\mu\nu}$ the Einstein tensor, and we assume that the affine connection vanishes in this coordinate system. The diagonal components of Eq.~\eqref{60.01_1} in STEGR become
\begin{equation}
E_{\lambda\lambda}=\hat{G}_{\lambda\lambda}=0\ ,\quad\text{with}\quad\lambda\in(t,r,\theta,\varphi)\ ,
\end{equation}
while the off-diagonal components become
\begin{equation}
E_{r\theta}=E_{\theta r}=-\frac{1}{2}f_{QQ}\frac{\partial Q}{\partial r}\cot\theta=0\ ,
\end{equation} 
that is, in  STEGR,  the solution~\eqref{60.01} satisfies Eqs.~\eqref{60.01_1}. Then, the Levi-Civita connection in these coordinates $\hat{\Gamma}^{\alpha}_{\phantom{\alpha}\mu\nu}$ for Schwarzschild metric, has the following non-vanishing components
\begin{align}\label{60.02}
&\hat{\Gamma}^{1}_{\phantom{1}00}=\frac{c^2r_{s}\left(r-r_{s}\right)}{2r^3}\ , &&\hat{\Gamma}^{0}_{\phantom{0}01}=-\hat{\Gamma}^{1}_{\phantom{1}11}=\frac{r_{s}}{2r\left(r-r_{s}\right)}\ , &&\hat{\Gamma}^{2}_{\phantom{2}12}=\hat{\Gamma}^{3}_{\phantom{3}13}=\frac{1}{r}\nonumber\ ,
\\&\hat{\Gamma}^{1}_{\phantom{1}22}=-\left(r-r_{s}\right)\ ,&&\hat{\Gamma}^{3}_{\phantom{3}23}=\cot\theta\ ,&& \hat{\Gamma}^{1}_{\phantom{1}33}=-\left(r-r_{s}\right)\sin^2\theta\ ,\nonumber\\
&\hat{\Gamma}^{2}_{\phantom{2}33}=-\sin\theta\cos\theta\ .
\end{align}
Taking into account that, in coincident gauge, the Levi-Civita connection is opposite to the disformation tensor and that the covariant derivative with respect to the affine connection reduces to the partial derivative, that is,
\begin{equation}
\hat{\Gamma}^{\alpha}_{\phantom{\alpha}\mu\nu}=-L^{\alpha}_{\phantom{\alpha}\mu\nu}\ ,
\end{equation}
\begin{equation}
\nabla_{\mu}=\partial_{\mu}\ ,
\end{equation}
the non-metricity scalar in Eq.~\eqref{19}, in coincident gauge, becomes
\begin{equation}\label{60.04}
Q=g^{\mu\nu}\Bigl(\hat{\Gamma}^{\alpha}_{\phantom{\alpha}\beta\mu}\hat{\Gamma}^{\beta}_{\phantom{\beta}\nu\alpha}-\hat{\Gamma}^{\beta}_{\phantom{\beta}\alpha\beta}\hat{\Gamma}^{\alpha}_{\phantom{\alpha}\mu\nu}\Bigr)\ .
\end{equation}
From the inverse diagonal element of  metric~\eqref{60.01} and the Levi-Civita connection~\eqref{60.02}, we obtain after a straightforward calculation, an expression for the scalar $Q$ in Eq.~\eqref{60.04}, that is
\begin{equation}\label{60.05}
Q=-\frac{2}{r^2}\ .
\end{equation}
In order to obtain  the result in Eq.~\eqref{60.05}, we used the following intermediate results
\begin{align}
g^{00}\Bigl(\hat{\Gamma}^{\alpha}_{\phantom{\alpha}\beta 0}\hat{\Gamma}^{\beta}_{\phantom{\beta}\alpha 0}-\hat{\Gamma}^{\alpha}_{\phantom{\alpha}\beta\alpha}\hat{\Gamma}^{\beta}_{\phantom{\beta}00}\Bigr)=\frac{r_{s}^2}{2r^3\left(r-r_{s}\right)}-\frac{r_{s}}{r^3}\ ,\\
g^{11}\Bigl(\hat{\Gamma}^{\alpha}_{\phantom{\alpha}\beta 1}\hat{\Gamma}^{\beta}_{\phantom{\beta}\alpha 1}-\hat{\Gamma}^{\alpha}_{\phantom{\alpha}1\alpha}\hat{\Gamma}^{1}_{\phantom{1}11}\Bigr)=-\frac{r_{s}^2}{2r^3\left(r-r_{s}\right)}-\frac{2}{r^2}+\frac{r_{s}}{r^3}\ ,\\
g^{22}\Bigl(\hat{\Gamma}^{\alpha}_{\phantom{\alpha}\beta 2}\hat{\Gamma}^{\beta}_{\phantom{\beta}\alpha 2}-\hat{\Gamma}^{\alpha}_{\phantom{\alpha}1\alpha}\hat{\Gamma}^{1}_{\phantom{1}22}\Bigr)=-\frac{\cot^2\theta}{r^2}\ ,\\
g^{33}\Bigl(\hat{\Gamma}^{\alpha}_{\phantom{\alpha}\beta 3}\hat{\Gamma}^{\beta}_{\phantom{\beta}\alpha 3}-\hat{\Gamma}^{\alpha}_{\phantom{\alpha}\beta\alpha}\hat{\Gamma}^{\beta}_{\phantom{\beta}33}\Bigr)=\frac{\cot^2\theta}{r^2}\ .
\end{align}
It  is worth noticing that although the non-metricity $Q$~\eqref{60.05} is a coordinate-independent scalar, it depends on the residual gauge, i.e., the $\xi^{\mu}$. So, since the Levi-Civita curvature scalar is vanishing for the Schwarzschild metric~\eqref{60.01},  nonzero $Q$ implies, via Eq.
\eqref{19_7},  that the boundary terms $B$ of GR and STEGR, in that particular $\xi$-gauge, differ. This means that their equivalence can be valid   at the level of field equations but not necessarily for all physical quantities. For a  discussion on this point, see e.g. Refs.~\cite{Gomes:2022vrc,Bahamonde:2022zgj}).
Finally, the time-time component of the gravitational pseudo-tensor~\eqref{30} for $f(Q)=Q$ is
\begin{equation}\label{60.06}
\tau^{0}_{\phantom{0}0|Q}=\frac{1}{2\kappa^{2}}\Bigl[P^{0}_{\phantom{0}\mu\nu}\partial_{0}g^{\mu\nu}+Q\delta^{0}_{0}-\frac{4\kappa^{2}}{\sqrt{-g}}\lambda_{\omega}^{\phantom{\omega}\beta 0 \nu}\,\partial_{0}\Gamma^{\omega}_{\phantom{\omega}\nu\beta}\Bigr]
\ .
\end{equation}
Taking into account that the metric~\eqref{60.01} is stationary, it is $\partial_{0}g^{\mu\nu}=0$. Hence it is  time independent, and, in coincident gauge, $\partial_{0}\Gamma^{\omega}_{\phantom{\omega}\nu\beta}=0$.  The gravitational energy density in  STEGR, for the Schwarzschild spacetime in spherical coordinates~\eqref{60.06}, becomes
\begin{equation}
\tau^{0}_{\phantom{0}0|Q}=\frac{Q}{2\kappa^2}\ .
\end{equation}
From Eq.~\eqref{60.05}, it  takes the  form
\begin{equation}\label{60.07}
\boxed{
\tau^{0}_{\phantom{0}0|Q}(r)=-\frac{Mc^2}{4\pi r_{s}}\frac{1}{r^2}}\ ,
\end{equation}
where we have taken into account the explicit expressions of the coupling constant $\kappa^2$.
Integrating the gravitational energy density~\eqref{60.07} in the spherical shell of inner radius $r_{s}$ and outer radius $R$, we have
\begin{equation}\label{60.08}
E_{g}(R)_{shell}=\int_{shell} d^{3}r\,\tau^{0}_{\phantom{0}0|Q}(r)=4\pi\int_{r_{s}}^{R}\, \tau^{0}_{\phantom{0}0|Q}(r)r^2dr\ ,
\end{equation}
that is
\begin{equation}\label{60.09}
\boxed{
E_{g}(R)_{shell}=Mc^2\Bigl(1-\frac{R}{r_{s}}\Bigr)}\ .
\end{equation}
If $R$ goes to infinity in Eq.~\eqref{60.09}, the gravitational energy diverges. This is due to the non-localizability of gravitational energy and thus the affine nature of the pseudotensor, reproducing the result of General Relativity obtained by Bauer (1918) in~\cite{Bau}. Similarly, the gravitational energy~\eqref{60.09} depends not only on the coordinates but also on the choice of the $\xi$'s.  So by choosing another symmetric teleparallel connection, the gravitational energy, associated with the Schwarzschild solution,  could be different. This will be the subject of forthcoming research.

\section{The weak field limit of  $f(Q)$ gravitational  pseudo-tensor}\label{sec8}
We now derive  the lower order perturbation of the $f(Q)$ energy-momentum pseudo-tensor  around the Minkowskian spacetime considering in coincidence gauge. The perturbed metric tensor reads as 
\begin{equation}\label{60}
g_{\mu\nu}=\eta_{\mu\nu}+h_{\mu\nu}\ ,
\end{equation}
with $\eta_{\mu\nu}$ the Minkowski metric  and $\vert h_{\mu\nu}\vert \ll 1$.

At the first order in metric perturbations, we get the following linear corrections, in coincident gauge, for non-metricity quantities coming from Eqs.~\eqref{5},\eqref{7} and \eqref{8}. It is
\begin{equation}\label{60.1}
Q_{\alpha\mu\nu}^{(1)}=\partial_{\alpha}h_{\mu\nu}\ ,
\end{equation}
\begin{equation}\label{60.2}
{Q_{\alpha}^{\phantom{\alpha}\mu\nu}}^{(1)}=\partial_{\alpha}h^{\mu\nu}\ ,
\end{equation}
\begin{equation}\label{60.2.1}
{Q}^{\alpha(1)}=\eta^{\alpha\beta}\partial_{\beta}h\ ,
\end{equation}
\begin{equation}\label{60.3}
{\widetilde{Q}}^{\alpha(1)}=\partial_{\beta}h^{\alpha\beta}\ ,
\end{equation}
where $h$ is the trace of the metric perturbation $h_{\mu\nu}$.
It follows that the linearized deformation tensor $L^{\alpha}_{\phantom{\alpha}\mu\nu}$,  defined in  Eq.\eqref{15}, becomes
\begin{equation}\label{60.4}
{L^{\alpha}_{\phantom{\alpha}\mu\nu}}^{(1)}=\frac{1}{2}\partial^{\alpha}h_{\mu\nu}-\partial_{(\mu}h^{\alpha}_{\phantom{\alpha}\nu)}\ ,
\end{equation}
while the linearized non-metricity conjugate tensor $P^{\alpha}_{\phantom{\alpha}\mu\nu}$ in Eq.\eqref{19_1} yields 
\begin{equation}\label{60.5}
{P^{\alpha}_{\phantom{\alpha}\mu\nu}}^{(1)}=-\frac{1}{4}\partial^{\alpha}h_{\mu\nu}+\frac{1}{2}\partial_{(\mu}h^{\alpha}_{\phantom{\alpha}\nu)}+\frac{1}{4}\bigl(\partial^{\alpha}h-\partial_{\beta}h^{\alpha\beta}\bigr)\eta_{\mu\nu}-\frac{1}{4}\delta^{\alpha}_{(\mu}\partial_{\nu)}h\ .
\end{equation}
Then if we expand $f(Q)$ up to the second order in $h$ as 
\begin{equation}\label{60.6}
f(Q)=f(0)+f_{Q}(0)Q+O(Q^2)\ ,
\end{equation} 
we get
\begin{equation}\label{60.7}
f(Q)^{(0)}=f(0)=0=f(Q)^{(1)}\ ,\quad\text{and}\quad f(Q)^{(2)}=f_{Q}(0)Q^{(2)}\ ,
\end{equation}
\begin{equation}\label{60.8}
f_{Q}^{(0)}=f_{Q}(0)\ ,\quad\text{and}\quad f_{Q}^{(1)}=0\ ,
\end{equation}
assuming $f(0)=0$, and $f(Q)^{(1)}=0$ because only second order terms in $h_{\mu\nu}$ remain.  On the other hand, the non-metricity scalar $Q$, defined in Eq.~\eqref{19}, up to the second order in $h$, from the Eqs.~\eqref{10},  \eqref{60.2} and \eqref{60.5}, becomes 
\begin{equation}\label{61}
Q^{(2)}=-\frac{1}{4}\partial^{\alpha}h_{\mu\nu}\partial_{\alpha}h^{\mu\nu}+\frac{1}{2}\partial_{\mu}h^{\alpha}_{\phantom{\alpha}\nu}\partial_{\alpha}h^{\mu\nu}+\frac{1}{4}\partial^{\alpha}h\,\partial_{\alpha}h-\frac{1}{2}\partial_{\alpha}h\,\partial_{\beta}h^{\alpha\beta}\ .
\end{equation}
Finally, from Eqs.~\eqref{60.2},\eqref{60.5}, \eqref{60.8}, and \eqref{61}, the second-order perturbations of the gravitational energy-momentum pseudo-tensor~\eqref{56}, in coincident gauge, becomes
\begin{equation}
\boxed{
\begin{aligned}
{\tau^{\alpha}_{\phantom{\alpha}\lambda}}^{(2)}=\frac{f_{Q}}{2\kappa^2}\Biggl[&-\frac{1}{4}\partial^{\alpha}h_{\mu\nu}\partial_{\lambda}h^{\mu\nu}+\frac{1}{2}\partial_{\mu}h^{\alpha}_{\phantom{\alpha}\nu}\partial_{\lambda}h^{\mu\nu}+\frac{1}{4}\partial^{\alpha}h\,\partial_{\lambda}h\\&-\frac{1}{4}\partial_{\beta}h^{\alpha\beta}\partial_{\lambda}h-\frac{1}{4}\partial_{\lambda}h^{\alpha\beta}\partial_{\beta}h\\&
+\Biggl(-\frac{1}{4}\partial^{\sigma}h_{\mu\nu}\partial_{\sigma}h^{\mu\nu}+\frac{1}{2}\partial_{\mu}h^{\sigma}_{\phantom{\sigma}\nu}\partial_{\sigma}h^{\mu\nu}+\frac{1}{4}\partial^{\alpha}h\,\partial_{\alpha}h-\frac{1}{2}\partial_{\sigma}h\,\partial_{\beta}h^{\sigma\beta}\Biggr)\delta^{\alpha}_{\lambda}\Biggr]
\end{aligned}
}\ .
\end{equation}
This lowest-order approximation of the gravitational energy-momentum pseudo-tensor  is  useful in order to  calculate the power and angular momentum carried by gravitational waves, as well as in deriving formulas that extend the gravitational quadrupole formula in non-metric gravitational theories \cite{CCN2,CC,CCL2}.

\section{Discussion and Conclusions}\label{sec9}
In this paper, we have derived the energy and momentum densities of the gravitational field in the  $f(Q)$ non-metric gravity, described by an affine tensor $\tau^{\alpha}_{\phantom{\alpha}\lambda}$ and not by a covariant one, i.e., the pseudo-tensor which is an  non-gauge invariant quantity. For this purpose, we have imposed  the invariance of only the gravitational part of the Lagrangian action under rigid translations. 

The  first local variation of the action without fixing the $x^{\alpha}$ coordinates, due to the stationary condition of the metric and connection field equations  in  vacuum,  allows us to derive the Noether currents that we  identify as a gravitational pseudo-tensor, and therefore a local continuity equation. Subsequently, from the vanishing of the LC covariant divergence of the energy-momentum tensor $T^{\alpha}_{\phantom{\alpha}\lambda}$, we have verified that, on-shell, the continuity equation holds for  both gravitational and matter contributions to the energy and momentum densities, i.e. the complex of energy and momentum contributions $\mathcal{T}^{\alpha}_{\phantom{\alpha}\lambda}$ is locally conserved in $f(Q)$ non-metric gravity.  In coincident gauge,  the pseudo-tensor can be perturbed up to the second order in $h_{\mu\nu}$, so one can  obtain a  useful expression  to calculate the power carried by a propagating gravitational wave as well as a gravitational multipole formula that  can contain additional terms with respect to the standard  quadrupole of GR.

We also pointed out the analogy between the pseudo-tensors in $f(T)$ and $f(Q)$ considering  the  Weitzenb\"ock connection and the coincident gauge.  Precisely, the role played by the inverse tetrad, the torsion conjugate and torsion scalar in $f(T)$  teleparallel gravity is played by the inverse of the metric tensor, the non-metricity conjugate and non-metricity scalar in $f(Q)$ gravity, respectively. 

As an application of the gravitational pseudotensor, we derived  the gravitational energy density  for a static and spherically symmetric spacetime, in the region outside the Schwarzschild radius, in the STEGR framework where $f(Q)$ reduces to $Q$ by adopting the concident gauge.

In a forthcoming study, these quantities will  be confronted with observations in view to find possible signature for $f(Q)$ non-metric gravity.

 \section*{Acknowledgements}
The authors acknowledge the Istituto Nazionale di Fisica Nucleare (INFN) Sez. di Napoli,  Iniziative Specifiche QGSKY and MoonLight-2  and the Istituto Nazionale di Alta Matematica (INdAM), gruppo GNFM, for the support.
This paper is based upon work from COST Action CA21136 -- Addressing observational tensions in cosmology with systematics and fundamental physics (CosmoVerse), supported by COST (European Cooperation in Science and Technology).

\end{document}